\documentclass[usenatbib,fleqn,useAMS]{mnras}
\usepackage{graphicx}
\usepackage{amsmath}
\usepackage{txfonts}
\usepackage{braket}

\usepackage{CJK}
\usepackage[utf8]{inputenc}

\newcommand\kname{\CJKfamily{mj}안진혁~(安振爀)}
\newcommand\kadd{\CJKfamily{mj}대전광역시~유성구~대덕대로~776, 한국천문연구원~이론천문센터}

\renewcommand\le\oldleq
\renewcommand\ge\oldgeq
\renewcommand\pi\upi
\newcommand\pdm\upartial
\newcommand\dm{\mathrm d}
\newcommand\im{\mathrm i}
\newcommand\rme{\mathrm e}
\newcommand\dirac\deltaup

\title[the central caustic degeneracy]
{On the condition for the central caustic degeneracy of
the planetary microlensing}

\author[An]
{J.~An\thanks{\kname:~E-mail:~{\tt jin(at)kasi.re.kr}}
\\Center for Theoretical Astronomy, Korea Astronomy \& Space Science Institute,
776 Daedeok-daero, Yuseong-gu, Daejeon 34055,\thanks{\kadd}
Republic of Korea (South).}

\pagerange{\pageref{firstpage}--\pageref{lastpage}}\volume{000}\pubyear{0000}
\date{Draft version 10 Feburary 2021}

\begin{document}
\label{firstpage}
\begin{CJK*}{UTF8}{}
\maketitle

\begin{abstract}
It is shown that the linear approximation of the central caustic
for the planetary ($q\ll1$) microlensing is valid if $|1-s|\gg q^{1/3}$
(where $q$ is the mass ratio and $s$ is the projected separation in the
unit of the Einstein ring radius of the primary). The condition is also
consistent with the requirement that the binary separation is
far from those in the resonant binary regime resulting in a single six-cusp
caustic. Given that the linear approximation of the caustic is invariant
under $s\leftrightarrow s^{-1}$, the close/wide binary degeneracy observed
under the same condition may be understood via the linear approximation
of the central caustic. Finally it is argued that the local degeneracies
of lensing features associated with caustic crossings can still persist
in the planetary events even when $|1-s|\sim q^{1/3}$ although
the overall caustic shape may not be degenerate at all.
\end{abstract}
\end{CJK*}

\section{Introduction}

Recently, \citet{Ye21} have reported a planetary ($q\sim10^{-5}$)
microlensing event with two degenerate solutions resembling
the classical $s\leftrightarrow s^{-1}$ degeneracy \citep{GS98,Do99}
but $s\sim1$. As the close/wide binary degeneracy is originally derived
from a certain symmetry present in the lens equation when the projected
separation is far from the Einstein ring radius, they have questioned
whether the observed degeneracy can be considered as a particular manifestation
of the well-known degeneracy. In fact, if the mass ratio is sufficiently
small so that the system can be approximated as a point mass lens under
a planetary perturbation, the central astroid caustic is also
invariant under the same $s\leftrightarrow s^{-1}$ transformation
of the separation up to the linear order of the perturbation \citep{Bo99,An05}.
Whilst this approximation is known to fail as $s\to1$,
the condition under which the approximation is valid is only vaguely mentioned
in the literature and so it is difficult to judge if the observed degeneracy
can be properly understood as an example of this classical approximation
approach. This short paper attempts to rectify the situation and address
more precise condition for the central caustic degeneracy of the planetary
microlensing.

\section{the caustics of the binary lens system}

Let us consider the gravitational lensing system described by
the complexified lens equation \citep{Wi90} given by
\begin{equation}\label{eq:leq}
\zeta=z-\overline{\alpha(z)};\quad
\alpha(z)=\frac{m_1}{z}+\frac{m_2}{z-\ell},
\end{equation}
where $\zeta$ and $z$ are the complexified source and image positions
and the overbar notation represents the complex conjugation. 
The system corresponds to the binary lens system with the primary of
the mass $m_1$ at the origin and the secondary of the mass $m_2$
at the location $\ell$. The Jacobian of the lens mapping is given by
\begin{equation}
J=1-\frac{\pdm\zeta}{\pdm\bar z}\frac{\pdm\bar\zeta}{\pdm z}
=1-\left\lvert\frac{\dm\alpha}{\dm z}\right\rvert^2,
\end{equation}
and the critical curve is defined to be the lens plane locus of $J=0$,
whereas the caustic is the source plane image of the critical curve
under the mapping in equation (\ref{eq:leq}).

The critical curve $z_{\rm cc}(\phi)$ may also be seen as
the parametric curve satisfying
$f(z_{\rm cc})=\rme^{-2\im\phi}$ \citep{Wi90} where
\begin{equation}\label{eq:cc}
f(z)=-\frac{\dm\alpha}{\dm z}=\frac{m_1}{z^2}+\frac{m_2}{(z-\ell)^2},
\end{equation}
and the caustic is similarly parametrized through
$\zeta_{\rm}(\phi)=z_{\rm cc}-\overline{\alpha(z_{\rm cc})}$.
Unless $f'(z_{\rm cc})=0$, the parametric critical curve
$z_{\rm cc}(\phi)$ is differentiable
such that $f'(z)\,z_{\rm cc}'(\phi)=-2\im\rme^{-2\im\phi}$,
and it follows that
\begin{equation}
\zeta_{\rm ca}'(\phi)=z_{\rm cc}'(\phi)
+\overline{f(z_{\rm cc})\,z_{\rm cc}'(\phi)}
=2\frac{\bar f'-f'\rme^{6\im\phi}}{\im|f'|^2\rme^{2\im\phi}}.
\end{equation}
Hence if $\bar f'=f'\rme^{6\im\phi}$ or equivalently
$f'^2\rme^{6\im\phi}=f'^2/f^3=|f'|^2\in\mathbb R^+$ is
positive real at a point on the critical curve, then
$\zeta_{\rm ca}'(\phi)=0$ at the corresponding point on the caustic
\citep{DH15}; that is, the parametrized caustic is locally stationary
and so it develops a cusp there. If $f'(z_{\rm cc})=0$ on the other hand,
it can be shown that $f(z_{\rm cc})=\rme^{-2\im\phi}$ possesses
a degenerate solution. This implies that the critical curve
(and the caustic as well) bifurcates at the corresponding
point and globally it becomes self-intersecting.

For the binary lens system, \citet{SW86} have shown that there exist
three possible topologies for the caustics depending on the projected
separation between two lens components: that is,
one astroid and two deltoids for ``close'' binaries,
two astroids for ``wide'' binaries, and a single simply-connected curve
with six cusps for intermediate (``resonant'') cases.
At the transition between these regimes,
the caustics and the critical curves become self-intersecting and so
the exact values of the separations dividing these three regimes correspond
to those permitting simultaneous solutions for
$f(z)=\rme^{-2\im\phi}$ and $f'(z)=0$. For $f(z)$ in equation (\ref{eq:cc}),
the equation $f'(z)=0$ reduces to $(1-\ell/z)^3=-m_2/m_1$
and so its solution is $z_0=\ell(1+\omega q^{1/3})^{-1}$ where
$\omega\in\set{1,\rme^{\pm2\im\pi/3}=(-1\pm\!\sqrt3)/2}$ is a cube root of unity and $q=m_2/m_1$.
In order for $z_0$ to be a solution of $f(z)=\rme^{-2\im\phi}$,
we then must have
\begin{equation}
\ell^2\rme^{-2\im\phi}=m_1(1+\omega q^{1/3})^3
=(m_1^{1/3}+\omega m_2^{1/3})^3.
\end{equation}
Hence the separations dividing these three regimes are
\begin{equation}\label{eq:lw}
|\ell_\mathrm w|=(m_1^{1/3}+m_2^{1/3})^{\frac32},
\end{equation}
and
\begin{equation}\label{eq:lc}
|\ell_\mathrm c|
=\left\lvert\frac{2m_1^{1/3}-m_2^{1/3}\pm\im\!\sqrt3m_2^{1/3}}2\right\rvert^{\frac32}
=(m_1^{2/3}-m_1^{1/3}m_2^{1/3}+m_2^{2/3})^{\frac34}.
\end{equation}
Here $|\ell_\mathrm w|\ge(m_1+m_2)^{1/2}\ge|\ell_\mathrm c|$
(equal only if $m_1m_2=0$) and so
$|\ell_\mathrm w|$ is the minimum separation for the wide binaries
(two caustics) whilst $|\ell_\mathrm c|$ is the maximum separation
for the close binaries (three caustics). We note that
the value for $\ell_\mathrm w$ is first derived by
\citet[eq.~18]{ES93} and also reproduced in \citet{Do99}.
However, both authors only provide with an implicit equation
for $\ell_\mathrm c$. In fact we find
$(m_1+m_2)^2-|\ell_\mathrm c|^4=3(m_1m_2)^{1/3}|\ell_\mathrm c|^{8/3}>0$
for $|\ell_\mathrm c|$ in equation (\ref{eq:lc}),
which is equivalent to \citet[eq.~17]{ES93} if $m_1+m_2=1$
and to \citet[eq.~57]{Do99} if $m_1^{1/2}d_\mathrm c=|\ell_\mathrm c|$
and $q=m_1/m_2$.

\section{The linear approximation for the central caustic
in the planetary lensing}

For a point lensing, the critical curve is basically identical
to the Einstein ring: that is, $z=m_1^{1/2}\rme^{\im\phi}$ is
the solution to equation (\ref{eq:cc}) if $m_2=0$.
If we reparametrize the critical curve as a deviation from
the Einstein ring, namely
$z_{\rm cr}=m_1^{1/2}(1+\epsilon)\,\rme^{\im\phi}$,
equation (\ref{eq:cc}) is then reducible to the equation
for the fractional deviation $\epsilon$ (which is complex):
\begin{equation}\label{eq:e6}
\frac1{(1+\epsilon)^2}
+\frac{q}{(1-s\,\rme^{-\im\phi}+\epsilon)^2}=1,
\end{equation}
where $q=m_2/m_1$ and $s=m_1^{-1/2}\ell$. Under the assumptions
that\footnote{$(1-|s|)^2\le|1-s\,\rme^{-\im\phi}|^2\le(1+|s|)^2$}
$|\epsilon|\ll1$ and $|\epsilon|\ll\bigl\lvert1-|s|\bigr\rvert$,
expanding the left-hand side of equation (\ref{eq:e6}) in a power series
of $\epsilon$ and equating up to its linear term results in
\begin{equation}
\epsilon\approx\frac q{2(1-s\,\rme^{-\im\phi})^2}
\left[1-\frac q{(1-s\,\rme^{-\im\phi})^3}\right]^{-2}.
\end{equation}
It follows that, if $q\ll\bigl\lvert1-|s|\bigr\rvert^3$,
the critical curve in the linear approximation of $q$ is given by
\begin{equation}\label{eq:zla}
z_{\rm cr}\simeq m_1^{1/2}\rme^{\im\phi}
\left[1+\frac q{2(1-s\,\rme^{-\im\phi})^2}+\mathcal O(q^2)\right],
\end{equation}
Here $|\epsilon|\sim q|1-s\,\rme^{-\im\phi}|^{-2}$ and so
the original assumptions (i.e.\ $|\epsilon|\ll1$ and
$|\epsilon|\ll\bigl\lvert1-|s|\bigr\rvert$)
also hold if $q\ll\bigl\lvert1-|s|\bigr\rvert^3$.
Note that equations (\ref{eq:zla}) with $\phi$ and $\phi+\pi$ are
both approximate solutions to the same equation
$f(z)=\rme^{-2\im\phi}=\rme^{-2\im(\phi+\pi)}$,
and so the perturbative solution in equation (\ref{eq:zla})
accounts for two of four solutions of $f(z)=\rme^{-2\im\phi}$.

The preceding derivation is somewhat heuristic but it makes
the condition for its validity
(viz.\ $q\ll\bigl\lvert1-|s|\bigr\rvert^3$) rather more clear.
Alternatively if the (uniformly) convergent power series for
$\epsilon=\sum_{k=1}^\infty\varepsilon_kq^k$ exist, the coefficients
$\varepsilon_k$ can be determined by inserting
the expression back to equation (\ref{eq:e6}) or the equivalent
(quartic on $\epsilon$) polynomial equation,
$(1-s\,\rme^{-\im\phi}+\epsilon)^2+q(1+\epsilon)^2
=(1+\epsilon)^2(1-s\,\rme^{-\im\phi}+\epsilon)^2$
and assembling terms with the same power on $q$: which results in
\begin{multline}\label{eq:esr}
\epsilon=\frac q{2(1-s\,\rme^{-\im\phi})^2}
-\frac{(1+3s\,\rme^{-\im\phi})}{8(1-s\,\rme^{-\im\phi})^5}q^2
\\+\frac{1+8s\,\rme^{-\im\phi}+5s^2\rme^{-2\im\phi}}
{16(1-s\,\rme^{-\im\phi})^8}q^3+\cdots,
\end{multline}
the linear term of which is indeed consistent with
equation (\ref{eq:zla}). Without knowing the general expression for
the arbitrary coefficient $\varepsilon_k$, it is difficult to 
assess if the power series in equation (\ref{eq:esr})
actually converges. Nevertheless equation (\ref{eq:esr}) suggests
that the ratio between the subsequent terms behaves like
$\bigl\lvert(\varepsilon_{k+1}q^{k+1})/(\varepsilon_kq^k)\bigr\vert\sim
q|1-s\,\rme^{-\im\phi}|^{-3}$, and so one expects
the convergence criterion to exist in a form of
$q\bigl\lvert1-|s|\bigr\rvert^{-3}<C$ with some constant $C$.

As for the caustics, with
$z_{\rm cr}=m_1^{1/2}(1+\epsilon)\,\rme^{\im\phi}$ and
$\bar z_{\rm cr}=m_1^{1/2}(1+\bar\epsilon)\,\rme^{-\im\phi}$,
equation (\ref{eq:leq}) leads to
\begin{equation}\label{eq:cla}
\zeta_{\rm ca}=m_1^{1/2}\left(1
+\epsilon-\frac1{1+\bar\epsilon}-\frac q{1-\bar s\,\rme^{\im\phi}+\bar\epsilon}
\right)\rme^{\im\phi}.
\end{equation} 
Here if $\epsilon$ is given by equation (\ref{eq:esr})
and $\bar\epsilon$ is its complex conjugate,
expanding equation (\ref{eq:cla}) in a power series on $q$ then results in
\begin{multline}\label{eq:claq}
\frac{\zeta_{\rm ca}}{m_1^{1/2}\rme^{\im\phi}}
\simeq\frac q2\left[\frac1{(1-s\,\rme^{-\im\phi})^2}
+\frac1{(1-\bar s^{-1}\rme^{-\im\phi})^2}-1\right]
\\-\frac{q^2}8\left(\frac{1+3s\,\rme^{-\im\phi}}{(1-s\,\rme^{-\im\phi})^5}
-\frac{5-\bar s^{-1}\rme^{-\im\phi}}
{\bar s^4\rme^{4\im\phi}(1-\bar s^{-1}\rme^{-\im\phi})^5}\right)+\cdots.
\end{multline}
Note that the linear term is invariant under the transform
$s\leftrightarrow\bar s^{-1}$, but this invariance does not hold
for the higher order term. Formally,
if we consider the relative departure from the invariance, namely
$\delta=(\tilde\zeta_{\rm ca}-\zeta_{\rm ca})/
(\tilde\zeta_{\rm ca}+\zeta_{\rm ca})$ where
$\tilde\zeta_{\rm ca}$ is equation (\ref{eq:claq}) under
the transformation of $s\to\bar s^{-1}$, we find that
$|\delta|\simeq q|\chi|^{-3}$ as $\chi\to0$ where
$\chi=1-s\,\rme^{-\im\phi}$. In other words, the central caustic
invariance under $s\leftrightarrow\bar s^{-1}$ is again applicable
(globally) when $\bigl\lvert1-|s|\bigr\rvert\gg q^{1/3}$.

\subsection{Relation to the non-resonance condition}

The linear-$q$ approximation of the caustic in equation (\ref{eq:claq})
results in an (distorted) astroid-shaped curve with four cusps, which
corresponds to the ``central caustics'' of either close or wide planetary
binaries. For the true binary lens system however, such caustics only exist
if the separation is not in the resonant region; that is,
\begin{equation}
|s|>m_1^{-1/2}|\ell_\mathrm w|=(1+q^{1/3})^{\frac32}
\simeq1+\frac32q^{1/3}+\frac38q^{2/3}+\cdots
\end{equation}
or
\begin{equation}\begin{split}
|s|<m_1^{-1/2}|\ell_\mathrm c|&=(1-q^{1/3}+q^{2/3})^{\frac34}
=\Biggl[\frac34+\biggl(\frac12-q^{1/3}\biggr)^2\Biggr]^{\frac34}
\\&\simeq1-\frac34q^{1/3}+\frac{21}{32}q^{2/3}+\cdots,
\end{split}\end{equation}
where $\ell_\mathrm w$ and $\ell_\mathrm c$ are the transition value
derived in equations (\ref{eq:lw}) and (\ref{eq:lc}).
With a sufficiently small $q$, this condition is
approximatly\footnote{The relation is \emph{not} actually valid
for $|s|<1$ if $q>0$. Rather the actual precise statement would be
$\lim_{q\to0}q^{1/3}/(1-s_\mathrm c)=4/3$ with
$s_\mathrm c=m_1^{-1/2}|\ell_\mathrm c|$. Alternatively,
the relevant zeroth order model for the close binary is the point mass
with the total mass of the system and
$q^{1/3}/[(1+q)^{1/2}-s_\mathrm c]$, albeit an increasing function of $q$,
is still bounded for all $q\in[0,1]$.} equivalent to \citep[cf.][eq.~62]{Do99}
\begin{equation}
\frac{q^{1/3}}{|s|-1}\la\frac23\quad
\text{for $|s|>1$, or }\
\frac{q^{1/3}}{1-|s|}\la\frac43\quad
\text{for $|s|<1$}.
\end{equation}
That is to say, the condition for the linear approximation to be valid
is more or less equivalent to the planetary system being far from
the resonant lensing regime. 

\begin{figure}
\centering\includegraphics[width=\hsize]{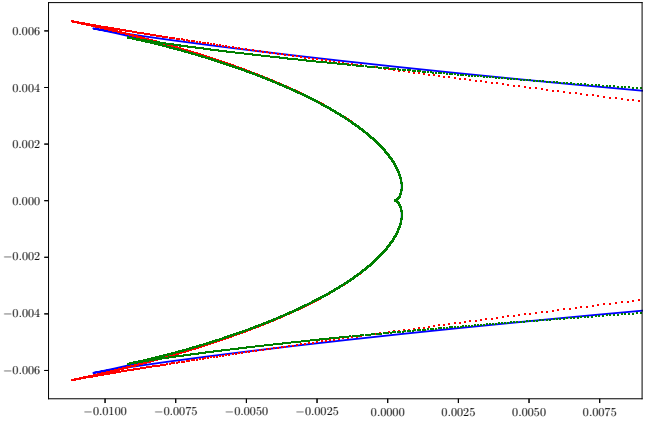}
\caption{\label{fig1}True and approximate caustics for
the planetary lens system with $q=10^{-3}$ near the primary location.
The red and green dots are respectively
for a close binary with $s=0.9$ and a wide binary with $s=0.9^{-1}$,
whilst the blue curve represents the approximation given by
the linear-$q$ terms in eq.~\ref{eq:claq}. Note that
$s_\mathrm c\approx0.9317(>0.9)$ and
$s_\mathrm w\approx1.1537(>0.9^{-1}\approx1.1111)$
for $q=10^{-3}$ and so the wide binary is actually in the resonant regime
and there is no discernible degeneracy in the global caustic shapes.
On the other hand, all three curves follow one another somewhat closely
in the region shown. The ``degeneracy'' is most severe closest to the primary
($s\rme^{-\im\phi}\sim-1$) and the curves become clearly distinguishable
around the next cusp locations although they are still similar.}
\end{figure}

\subsection{Global vs local degeneracy}

\begin{figure}
\centering\includegraphics[width=\hsize]{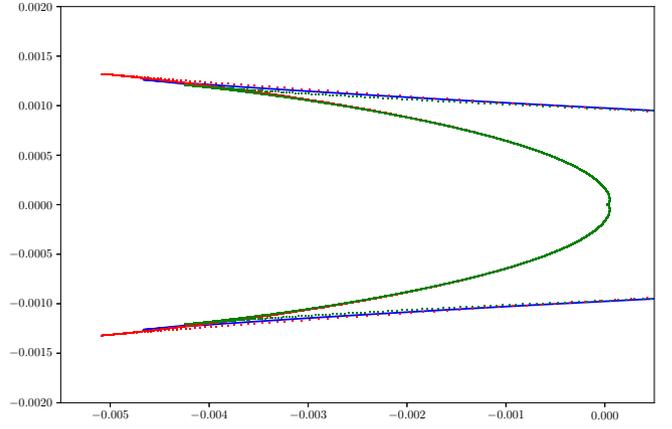}
\caption{\label{fig2}Same as Fig.~\ref{fig1} but $q=10^{-4}$
and $s=0.95$ or $0.95^{-1}$. Again $s_\mathrm c\approx0.9666(>0.95)$ and
$s_\mathrm w\approx1.0704(>0.95^{-1}\approx1.0526)$ and so there should be
no global degeneracy but the curves are quite similar locally.}
\end{figure}

Technically the preceding linear approximations of the critical curves
and the caustics start to break down as $s\rme^{-\im\phi}\to1$
and not in fact $|s|\to1$. In other words,
even if $\bigl\lvert1-|s|\bigr\rvert\la q^{1/3}$, the approximation
may be locally valid for the portion of the caustic corresponding to
the values of $\phi$ with which $s\rme^{-\im\phi}$ is sufficiently
far from the unity. For example, Figures~\ref{fig1} and \ref{fig2}
show the portions of caustics for the planetary
($q=10^{-3}$ and $10^{-4}$ respectively) lensing models with
$q^{1/3}/\bigl\lvert1-|s|\bigr\rvert\simeq1$. Whilst the approximation
in equation (\ref{eq:claq}) is invalid globally for these models
and the overall caustic shapes are completely different between
two models related by $s\leftrightarrow s^{-1}$, the caustic shown
in the region of Figures~\ref{fig1} and \ref{fig2} actually passably
resembles each other and also the linear approximation. In fact, we expect
that slight adjustments of $(q,s)$ separately for the close and wide binary
might result in a pair of locally degenerate models although they would not
be in the exact $s\leftrightarrow s^{-1}$ correspondence.

Whilst the true observed degeneracy of the microlensing lightcurve should
be properly analyzed in the magnification map and not just with the shape
of the caustics, most noticeable detailed structures of the microlensing
events are dominated by the caustic crossings and the cusp approaches.
Since the behaviour of the magnifications near the caustics is known to be
fairly generic \citep[e.g.,][]{KGP03,KGP05}, the resemblance of the caustics
may be a good starting point to understand the degeneracy in
the lightcurve modelling, especially given the fact that the lightcurve
actually samples the magnification map only along a 1-d slice.

\label{lastpage}
\end{document}